\documentclass[twocolumn,showpacs,preprintnumbers,amsmath,amssymb]{revtex4}
\usepackage{bm,amsmath,amssymb,latexsym,graphicx,enumerate}
\usepackage[mathcal]{euscript}
\usepackage{epsfig}

\newcommand{\be}{\begin{equation}}
\newcommand{\ee}{\end{equation}}

\begin{document}

\title{Fast control of the reflection of a ferroelectric
by an extremely short pulse}

\author{J.-G.~Caputo$^{1}$, A.I.~Maimistov$^{2,3}$  and E.V. Kazantseva$^{2,4}$ }
\affiliation{\normalsize \noindent $^1$:
Laboratoire de Math\'ematiques, INSA de Rouen, \\
Avenue de l'Universite, Saint-Etienne du Rouvray, 76801 France \\
$^2$: Department of Solid State Physics and Nanostructures, \\
Moscow Engineering Physics Institute, Kashirskoe sh. 31,
Moscow, 115409 Russia \\
$^3$: Department of General Physics,\\
Moscow Institute for Physics and Technology, Institutskii lane 9,
Dolgoprudny, Moscow region, 141700 Russia \\
$^4$: Department of Condensed Matter Physics, Moscow Institute of
Radiotechnics,\\
 Electronics and Automation, Vernadskogo pr. 78, Moscow,
119454 Russia \\
E-mails: elena.kazantseva@gmail.com, caputo@insa-rouen.fr,
aimaimistov@gmail.com \\
}
\date{\today}

\begin{abstract}
We propose a new type of optical switch based on a ferroelectric. 
It is based on the gap which exists for waves propagating 
from a dielectric to a ferroelectric material. This gap 
depends on the polarization of the ferroelectric.
We show that it can be shifted by a control electromagnetic pulse 
so that the material becomes transparent. 
This device would shift 
much faster than the relaxation time of the ferroelectric (1 nano s). 
Estimates are given for a real material.
\end{abstract}

\pacs{Ferroelectric materials, 77.84.-s ,Switching in ferroelectrics, 77.80.Fm
Optical switches, 42.79.Ta}

\maketitle

\section{Introduction}

The rapid control of light is a problem that has been studied for
a long time, in particular because of the applications. 
Ferroelectric materials are good candidates for this control because they can
be activated using an electric field. The principle is the following.
The reflection and transmission properties of a ferroelectric
material depend on its state of polarization, in particular it's
spontaneous polarization which exists in the absence of electric
field. The controlling electric field will shift this polarization 
so that then we can control the
reflection and transmission of any electromagnetic wave.

A first idea is to use a constant electric field such as in
\cite{Mishina:03}. However then we need to take into account the
relaxation of the ferroelectric which is about one nanosecond 
and this limits considerably the possibilities for applications.
Another idea is to use a fast electromagnetic pulse so that the
relaxation of the ferroelectric can be neglected. 
Here there are two
frequency regions, one is the high frequency signal we want to control,
the other, of lower frequency is the control
signal. To fix ideas we choose the high frequency signal to be about
a femtosecond in period and the control signal ten times slower. The
time scales are such that we can neglect the relaxation of the
polarization. This way we achieve light control with light, using an
ultra-short pulse.
In this article, we show that the
reflection coefficient is equal to one in a frequency region, a "gap"
whose position depends on the state of the polarization and the control.
Shifting the control, we move the gap and render the ferroelectric opaque
or transparent to femtosecond waves.\\
After deriving the model equations for the field and polarization, we linearize
them around a functioning point and derive the reflection and transmission
coefficients using a scattering formalism. These are then discussed as
a function of the control to show how light can be driven by the femtosecond
pulse.

\section{Basic equations}

The Maxwell equations in a dielectric medium are taken in the following form
\begin{eqnarray*}
\nabla\times\mathbf{E} &  =-\mathbf{B}_{,t},\qquad\nabla\cdot\mathbf{E}=-\epsilon_0^{-1}\nabla\cdot \mathbf{P}
,\\
c^2 \nabla\times\mathbf{B} & = \mathbf{E}_{,t}+\epsilon_0^{-1}
\mathbf{P}_{,t},\qquad \nabla\cdot\mathbf{B}=0,
\end{eqnarray*}
where the subscripts indicate the time derivative. From
these equation the wave equation results in
\[c^2\nabla^2 \mathbf{E}-\mathbf{E}_{,tt} =
\epsilon_0^{-1} [\mathbf{P}_{,tt} -\nabla(\nabla\cdot \mathbf{P})].
\]

\begin{figure}
\centerline{\epsfig{file=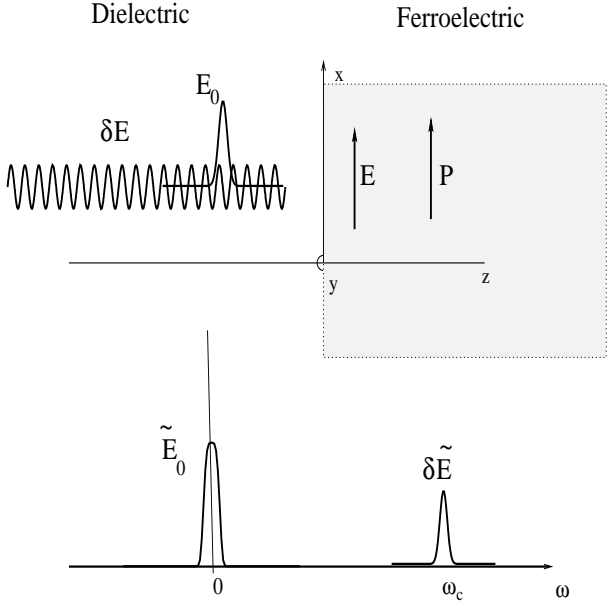,height=8 cm,width=8 cm,angle=0}}
\caption{
Top panel: schematic drawing of the dielectric-ferroelectric interface. The electric field $E$
is incident normally to the interface. It
is polarized along $x$ and the polarization is also along $x$. We show the control
pulse $E_0$ and the controlled small amplitude wave $\delta E$ of frequency $\omega_c$,
the carrier frequency
(see text for details). The
bottom panel indicates the Fourier spectra of these waves.
}
\label{f1}
\end{figure}

The geometry is shown in Fig. \ref{f1}, it is a dielectric layer
for $z<0$ and a ferroelectric layer for $z \ge 0$. We assume the
electric field to be polarized along $x$. The spontaneous polarization
of the ferroelectric is supposed to be also along $x$ and to depend only on
$z$. Because of this the wave equation above reduces to
\be\label{wave}
c^2 {E}_{,zz} - {E}_{,tt} = {{P}_{,tt} \over \epsilon_0}.
\ee
In particular the term $\nabla(\nabla\cdot \mathbf{P})$ is zero because the
polarization does not vary along $z$, the direction of propagation of
the pulse. The linear dielectric material ($z<0$) will be described by the Drude-Lorentz
model \cite{Rosenfeld}
so that the polarization $P$ follows
\be\label{drude_lorentz}P_{tt} + \omega_0^2 P = \epsilon_0 \omega_p^2 E,\ee
where $\omega_0$ is the polarization frequency and $\omega_p$ the plasma
frequency.
The polarization in the ferroelectric is given by the Landau-Kalatnikov
equation \cite{Landau_EDCM}
\be\label{landau_kala} \tau^2 P_{,tt} -A P + B P^3 = \epsilon_0 E,\ee
where $A = \alpha (T_c-T) >0$ and $B>0$ are the Landau-Ginzburg coefficients and
where $\tau$ is the characteristic time of the polarization. The three
equations (\ref{wave},\ref{drude_lorentz},\ref{landau_kala}) describe
completely the field and polarization of the media.

For zero electric field, the spontaneous polarization of the ferroelectric
is given by
\be\label{spont_polar}
-A P_0 + B P_0^3 =0,~~\to ~~P_0 = \pm \sqrt{A\over B}.\ee
Here the $+$ (resp. $-$) sign is for a polarization in the direction of $+x$ (resp. $-x$).
Now assume a field $E_0$ constant during a time interval $t_p$. The polarization
will then shift and satisfy
\be\label{polar_e0}
-A P_0 + B P_0^3 = \epsilon_0 E_0 .\ee
For small $E_0$ we can estimate $P_0$ using $E_0$ as a perturbation. We get
\be\label{p0e0}
P_0 = \pm \sqrt{A\over B} + {\epsilon_0 E_0 \over 2 A}
+ O(E_0^2) .\ee
Note that we did not consider the polarization close to 0 because it is
unstable.
During the time interval $t_p$ we can send a small
electromagnetic wave $\delta E$. This will shift the polarization by $\delta P$.
Assuming
$E = E_0 +\delta E,~~P = P_0 + \delta P$ where $|\delta E | <<  E_0$
$|\delta P| << P_0$ in equations
(\ref{wave},\ref{drude_lorentz},\ref{landau_kala}) yields the
linear system in the ferroelectric
\begin{eqnarray}\label{lin_ferro}
c^2 {\delta E}_{,zz} - {\delta E}_{,tt} = \epsilon_0^{-1}\delta P_{,tt} ,\\
\tau^2 \delta P_{,tt} -A \delta P + 3 B P_0^2 \delta P  = \epsilon_0 \delta E.
\end{eqnarray}
In the dielectric layer $P_0=0$ so that the second equation should
be replaced by
\be\label{lin_dielectric} \delta P_{,tt} + \omega_0^2 \delta P = \epsilon_0
\omega_p^2 \delta E .\ee
The three linear equations (\ref{lin_ferro},\ref{lin_dielectric}) represent
the small oscillations $ (\delta E, \delta P)$ around the functioning
point $(E_0,P_0)$ which exists during the time $t_p$.

\section{Solutions of the wave equations in frequency domain}

The system of linear equations above can now be solved completely using Fourier
transforms in $z$ and matching ${\delta E}$ and its derivative at the
interface $z=0$.
Taking the Fourier transform in $z$ we get for $z<0$
\begin{eqnarray}\label{epfourier_zm}
\tilde{\delta E}_{,zz} + k_0^{2}\tilde{\delta E} = -k_0^{2}\varepsilon_0^{-1}
\tilde{\delta P},  \\
\tilde{\delta P} = \varepsilon_0\tilde{E}
{\omega_p^2 \over \omega_0^2-\omega^2}.
\end{eqnarray}
Plugging the second equation into the first one, we obtain
\begin{equation}\label{efourier_zm}
\tilde{\delta E}_{,zz} +
k_0^{2}\left(1+{\omega_p^2 \over \omega_0^2-\omega^2}
\right)\tilde{\delta E} =0.
\end{equation}
In the ferroelectric medium for $z>0$,
starting from equations (\ref{lin_ferro}) and following a similar procedure as for $z<0$ we get
\be \label{efourier_zp2}
\tilde{\delta E}_{,zz} +
k_0^{2}\left(1+{1 \over -\tau^2 \omega^2-A + 3B P_0^2}
\right)\tilde{\delta E} =0.
\ee
Substituting the expression (\ref{p0e0}) of the spontaneous polarization $P_0$ we finally get
\be \label{efourier_zp}
\tilde{\delta E}_{,zz} +
k_0^{2}\left(1+{1 \over 2A \pm 3 \epsilon_0 E_0 \sqrt{B/A}
-\tau^2 \omega^2}
\right)\tilde{\delta E} =0.
\ee

Thus we have a piecewise wave equation for $z<0$ (\ref{efourier_zm}) and
$z>0$ (\ref{efourier_zp}).
In such linear media we can introduce the dielectric permittivity to
describe wave propagation. We get
\[
\varepsilon_{diel}(\omega)=
1+{\omega_p^2 \over \omega_0^2-\omega^2}
\] in the dielectric layer and
\[
\varepsilon_{ferr}(\omega)=
1+{1 \over 2A \pm 3 \epsilon_0 E_0 \sqrt{B/A}
-\tau^2 \omega^2}
\]
in the ferroelectric layer. The solution in the two different regions is
$$E(z,t) = e^{i\omega t -ik_1 z} ~~z<0, E(z,t) = e^{i\omega t -ik_2 z} ~~z>0 $$
where the wave numbers $k_1$ and $k_2$ are
\begin{eqnarray} \label{k1k2}
  && k_1 = k_0\left(
1+{\omega_p^2 \over \omega_0^2-\omega^2}
\right )^{1/2}, \\
  && k_2 = k_0\left(
1+{1 \over 2A \pm 3 \epsilon_0 E_0 \sqrt{B/A}
-\tau^2 \omega^2}
\right)^{1/2}.
\end{eqnarray}
These formulas show us that the external electromagnetic pulse can control
the dielectric properties of the ferroelectric material.

\section{Scattering of linear waves off the interface $z=0$.}

The reflection and transmission coefficients of harmonic waves can be computed
as a function of the control field $E_0$. Note that the two orientations
of polarization will give the same reflection coefficient for $E_0=0$. Only adding the
control $E_0$ is one able to distinguish the two states of polarization.
We set up the scattering formalism assuming an incident wave from the left, a
reflected wave and a transmitted wave,
$$\tilde{\delta E}(z,\omega)= E_{in}(\omega)e^{ik_1 z}+E_{r}(\omega)e^{-ik_1
z}, $$ in the dielectric and
$$ \tilde{E}(z,\omega)= E_{t}(\omega)e^{ik_2 z}, $$
in the ferroelectric medium.
In the absence of the surface charges and currents, the jump
conditions on the interface read \cite{Born:Wolf}
\[
\tilde{E}(0-,\omega)= \tilde{E}(0+,\omega), \quad
\tilde{E}_{,z}(0-,\omega)= \tilde{E}_{,z}(0+,\omega).
\]
Using these conditions and the solution of the wave equation
one can find the Fresnel relations connecting the amplitudes of the
incident wave $E_{in}(\omega)$, reflected wave $E_{r}(\omega)$ and
transmitted wave $E_{t}(\omega)$.
\begin{eqnarray}
E_{r}(\omega)&=& \frac{k_1-k_2}{k_1+k_2}E_{in}(\omega),  \label{fresn:1} \\
E_{t}(\omega)&=& \frac{2k_1}{k_1+k_2} E_{in}(\omega).\label{fresn:2}
\end{eqnarray}
These relations are correct for any low amplitude waves, both
solitary waves and for harmonic waves.
The reflection and transmission coefficients are then respectively
\be\label{ref_trans}
R = {E_{r} \over E_{in}} = \frac{k_1-k_2}{k_1+k_2},~~~~
T = {E_{t} \over E_{in}} = \frac{2k_1}{k_1+k_2},\ee
\begin{figure}
\centerline{\epsfig{file=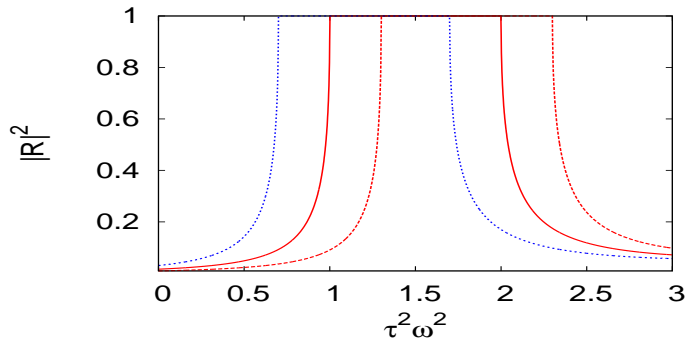,height=5cm,width=10 cm,angle=0}}
\caption{
Square of the modulus of the reflection coefficient
$|R|^2$ as a function of the reduced frequency $\tau^2 \omega^2$ for three different values
of the normalized control $3 \epsilon_0 E_0/P_s =0$ (continuous line, red online), $0.3$ (long
dash, red online) and $-0.3$ (short dash, blue online). The parameters are
$A=1,~~B=1$.
}
\label{f2}
\end{figure}

\begin{figure}
\centerline{\epsfig{file=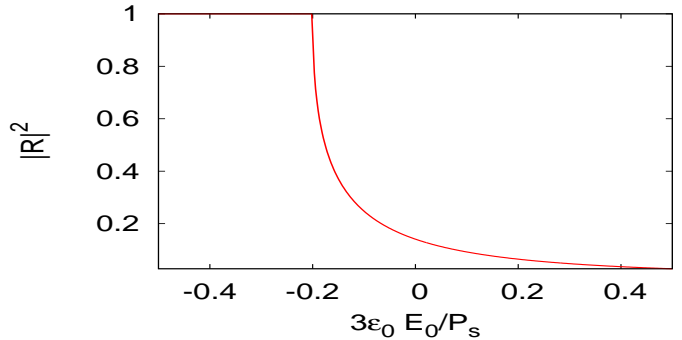,height=5cm,width=10 cm,angle=0}}
\caption{
Square of the modulus of the reflection coefficient
$|R|^2$ for a fixed reduced frequency $\tau^2 \omega^2 =0.8$ as a function
of the normalized control $3 \epsilon_0 E_0/ P_s$.
}
\label{f3}
\end{figure}

\section{Discussion}

Fig. \ref{f2} shows the modulus of the reflection coefficient $|R|^2$
as a function of the reduced frequency $\tau^2 \omega^2$ for three
different values of the control
$E_0=0$ (continuous line, red online), $E_0=0.3$ (long
dash, red online) and $E_0=-0.3$ (short dash, blue online). As can be seen
the boundary of the gap for which there is total reflection of the wave is shifted
to higher frequencies (resp. lower frequencies) for $E_0>0$  (resp. $E_0<0$).
We have assumed the + sign in the expression of $k_2$ (\ref{k1k2}).
To illustrate how the field $E_0$ can be used to block a wave we have plotted in Fig. \ref{f3}
$|R|^2$ as a function of $E_0$ for $\tau^2 \omega^2=0.8$. For $E_0 >0.4$
the ferroelectric is transparent. As $E_0$ is decreased $|R|^2$ increases sharply
and reaches 1 for $E_0=0.2$. Below that value the ferroelectric is opaque to
this particular frequency.

To show how this scheme can work in reality, we examine parameters
for a real material. Consider the study by Noguchi et al
\cite{Noguchi00} on the $Bi_4 Ti_3 O_{12}$­$Sr Bi_4 Ti_4 O_{15}$ intergrowth
ceramics. This material was shown to have a large spontaneous polarization. In
addition its Curie temperature is high so that it is stable. 
To estimate $A$ and $B$ from the measurements of \cite{Noguchi00} we recall that
$${A \over B}=P_c^2 ,~~~{2\over 3} A \sqrt{A \over 3 B}= \epsilon_0 E_c,$$
where $P_c$ and $E_c$ are respectively the coercitive polarization and coercitive field. Using these values from \cite{Noguchi00} we get
$$A=2 10^{-3},~~B = 10^{-1} m^4 C^{-2}.$$
This value of $A$ defines a resonant frequency $\omega_r$ such that
$$\tau^2 \omega_r^2 = 2 A.$$
We get $$\tau \omega_r =4.5 10^{-2}$$
The value of $\tau $ given by estimates of the inertia of molecular
assemblies is about $\tau = 10^{-10}$ \cite{cmmk10}.
This gives $$\omega_r = 4.5 10^8 Hz.$$
An important point is that at resonance the two terms $2 A$ and
$\tau^2 \omega^2$
are almost equal so their difference is very
small. Then a small shift due to the term
$3 \epsilon_0 E_0 \sqrt{B/A} $ will displace the resonance.
Let us estimate the field $E_0$ needed to shift the resonance from
$\omega_r$ to $\omega_r/2$. We have
$$3\epsilon_0 E_0 \sqrt{B \over A} = \tau^2 {\omega_r^2 \over 4}.$$
This gives
$$E_0 \approx 10^6 V m^{-1}.$$
This value of the electric field can be achieved using a laser.

\section{Conclusion}

The reflection of the electromagnetic wave on the interface between
linear dielectric medium and ferroelectric was considered. We assume
that the electromagnetic wave is a superposition of a high frequency wave
and a spike-like electromagnetic signal. The spectrum of the spike is
located near the zero frequency and can be considered as low frequency.
We showed that the spike-like signal induced an extra contribution
to the total ferroelectric polarization. This causes a fast change of
the reflection coefficient in the high frequency domain. In the time
duration of the spike-like signal the relaxation processes can be
neglected. Thus one can achieve a rapid control of light with light,
using an extremely short, spike pulse.


\begin{thebibliography}{999}


\bibitem{Mishina:03} E.D. Mishina, N.E. Sherstyuk, V.I. Stadnichuk, A. S. Sigov, V. M.
Mukhorotov, Yu. I. Golovko, A. van Etteger and Th. Rasing, Appl.
Phys. Lett. \textbf{83}, 12 2402-2404 (2003)


\bibitem{Rosenfeld} L. Rosenfeld, Theory of Electrons, New York: Dover Publications,
1965, pp. 68

\bibitem{Landau_EDCM} L.D. Landau, L.P. Pitaevskii, E.M. Lifshitz, \textit{Electrodynamics of
Continuous Media}, Elsevier Butterworth- Heinemann, Oxford, (2000).

\bibitem{Born:Wolf} M.Born, E. Wolf, Principles of Optics: Electromagnetic Theory of Propagation,
Interference and Diffraction of Light, seventh (expanded) ed.,
Cambridge Univ. Press, Cambridge, UK, 2003.

\bibitem{Ginzburg45} V.L. Ginzburg, JETP \textbf{15}, 739 (1945) (in Russian)

\bibitem{Devonshire} A.F. Devonshire Philos. Mag. \textbf{40} 1040 (1949)

\bibitem{Ginzburgufn} V.L. Ginzburg Phys. Usp. \textbf{44} 1037 (2001)

\bibitem{Noguchi00} Y. Noguchi, M.  Miyayama and T. Kudo, Appl. Phys. Lett.,
\textbf{77}, 3639-3641,  (2000).

\bibitem{cmmk10} J.-G.~Caputo, A.I.~Maimistov, E.D. Mishina, E.V. Kazantseva,  V.M.~Mukhortov,\\
Phys. Rev. B \textbf{82}, 094113, (2010).


\end{thebibliography}
\end{document}